\documentstyle[sprocl,epsfig]{article}

\def\haf{\textstyle{1\over2}}

\newskip\humongous \humongous=0pt plus 1000pt minus 1000pt
\def\caja{\mathsurround=0pt}

\newif\ifdtup
\def\panorama{\global\dtuptrue \openup1\jot \caja
        \everycr{\noalign{\ifdtup \global\dtupfalse
        \vskip-\lineskiplimit \vskip\normallineskiplimit
        \else \penalty\interdisplaylinepenalty \fi}}}
\def\eqalignno#1{\panorama \tabskip=\humongous
        \halign to\displaywidth{\hfil$\displaystyle{##}$
        \tabskip=0pt&$\displaystyle{{}##}$\hfil
        \tabskip=\humongous&\llap{$##$}\tabskip=0pt
        \crcr#1\crcr}}

\begin{document}

\title{NUCLEAR SCALES}
\author{J. L. FRIAR}

\address{Theoretical Division,Los Alamos National Laboratory,
Los Alamos, NM 87545 USA\\E-mail: friar@sue.lanl.gov} 

\maketitle\abstracts{ Nuclear scales are discussed from the nuclear physics
viewpoint. The conventional nuclear potential is characterized as a black box 
that interpolates nucleon-nucleon (NN) data, while being constrained by the 
best possible theoretical input. The latter consists of the longer-range parts
of the NN force (e.g., OPEP, TPEP, the $\pi$-$\gamma$ force), which can be
calculated using chiral perturbation theory and gauged using modern phase-shift 
analyses. The shorter-range parts of the force are effectively parameterized by 
moments of the interaction that are independent of the details of the force 
model, in analogy to chiral perturbation theory. Results of GFMC calculations in
light nuclei are interpreted in terms of fundamental scales, which are in good
agreement with expectations from chiral effective field theories. Problems with
spin-orbit-type observables are noted.}

\section{Introduction}

When I looked at the list of participants at this workshop, I realized that
nuclear physicists were in the minority (in spite of the workshop title). 
Because I have nothing to say about the $^1$S$_0$ partial wave of the
nucleon-nucleon (NN) system, I thought that it might be a reasonable idea to
discuss topics related to nuclear scales, but from the nuclear physics
viewpoint.  I will adopt as my purview those scales of nuclear physics that
somehow reflect the underlying dynamics.  I will examine in roughly equal
measure the two- and few-nucleon systems, and say a few words at the end about
the many-nucleon problem, where one critical test of naturalness has been made. 
Ultimately I hope to demonstrate that power counting based on the properties of
chiral effective field theories of nucleons and pions is manifested in nuclear
properties. This will be developed using a ``hybrid'' approach, where 
field-theoretic objects are imbedded in a traditional nuclear potential.
Beyond that, this talk can be viewed as a ``cultural'' exercise.

The few-nucleon systems are arguably the area of nuclear physics that has shown
the greatest progress in the past 15 years~\cite{joe,walter}. Although the
two-nucleon problem was solved decades ago, a convincing solution of the triton
($^3$H) problem at the 1\% level was not achieved until 1984~\cite{h3}.  In
fairly rapid succession (at roughly the same level of accuracy) we have added
$^3$He (the Coulomb interaction was initially a problem~\cite{he3}), the 
$\alpha$ particle~\cite{he4} ($^4$He), $^5$He (two $n$-$\alpha$
resonances~\cite{he5}), and the ground and low-lying excited states of A = 6, 7,
and 8 nuclei~\cite{a6-8}.  Scattering of neutrons~\cite{walter} and
protons~\cite{pisa} on deuterium (n - d and p - d) has become a sophisticated
industry.  Reactions (such as radiative capture) have also been
treated~\cite{reac}.

This area of nuclear physics has always been separate and has had one primary
{\it raison d'etre}:  studying the nuclear force.  The {\it modus operandi} has
been sophisticated computational methods (such as the Faddeev 
equations~\cite{walter} and Green's Function Monte Carlo [GFMC] 
techniques~\cite{joe}) together with brute-force
solutions using the biggest and fastest computers.  The key computational goal
in all of this is attaining 1\% accuracy in observables, which is typically good
enough.

\section{The Nuclear Potential}

Why do we do this?  Nuclear physics has a predictive theory predicated largely
on potential-based dynamics.  Given a nuclear potential, V, we can incorporate
it into a Hamiltonian and solve the Schr\"{o}dinger equation (or its equivalent,
such as the Faddeev equations).  A single (form for) V can lead to hundreds of
predictions that can be compared to data.  In addition, the tools developed for
unraveling nuclear dynamics (such as phase-shift analysis [PSA]) are now so
sensitive that they are capable of ``measuring'' various building blocks in the
dynamics that are of interest to both the nuclear- and particle-physics
communities~\cite{vanc}.  Finally, with the ability to perform detailed
calculations comes the capability of determining how the underlying scales are
reflected in nuclear properties~\cite{scales}.

Before we take a detailed look at these topics, a few words of history are in
order concerning why it took nuclear physics so long to get into this position. 
Unlike atoms, where the dominant (by far) interaction is a single type of
central force, the nuclear force is much more complicated.  This complexity can
be gauged by noting that two interacting nucleons with spin-$\haf$ and
isospin-$\haf$ have 4 possible spin values and 4 possible isospin values, for a
total of 16 possible spin-isospin components.  This is reflected in the
new~\cite{AV18} AV$_{18}$ and old~\cite{AV14} AV$_{14}$ potentials, which have 
18 and 14 components, respectively.  Moreover, the dominant type of force in a
nucleus is not central, but tensor in nature (like magnetic forces), and
noncentral forces make a theorist's life much, much harder.  The ability to
handle this complexity depends almost entirely on very large computational
resources, and this came rather late (in my career, at least).

Although the use of potentials is conventional in nuclear theory, a reasonable
question to ask is:  why should anyone take them seriously? There are quite a
few parameters in the potential (roughly 3 per spin-isospin component, which
are fit to NN data) and the amount of physics present in the interior region of
various force models ranges from modest to none.

To me the conventional nuclear potential is a ``black box'', an interpolating
mechanism whose input mixes the best available theory with NN scattering data. 
The output is a set of predictions of nuclear properties, as indicated by the
cartoon in Fig.\ 1.  To the extent that the mass of predictions outweighs the
number of parameters (it does), the procedure is predictive.  To a large extent
this workshop is predicated on substituting Chiral Perturbation Theory
($\chi$PT) for the traditional potential in the black box.

\begin{figure}[htbp]
  \epsfig{file=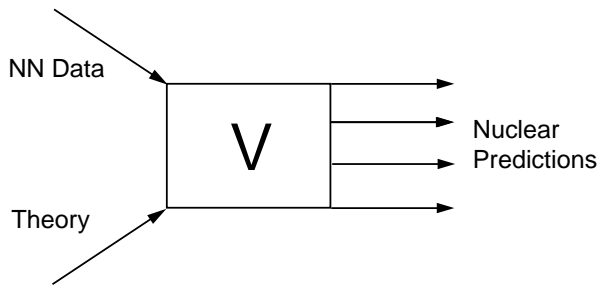,height=1.50in,bbllx=65pt,bblly=390pt,bburx=483pt,
   bbury=550pt}
  \caption{The nuclear potential as a black box that interpolates NN
  data through the filter of the best possible nuclear theory and effectuates
  predictions of nuclear properties.}
\end{figure}

There are actually two closely-related black boxes.  One is the
phase-shift-analysis black box, and the other is the nuclear-potential black
box.  In the distant past they were not so closely related.  Those who did PSA 
were typically not those who developed potential models, and the latter were
usually not fit directly to the NN data base as they are now.  Thus, the
``first-generation'' potentials were not especially good fits to the NN data
base.  Today's ``second-generation'' potentials~\cite{AV18,pots} range from good
to excellent fits. In other words, the ``black box'' parameterizes very well.

The typical strong-interaction physics input is illustrated in Fig.\ (2).  The
longest-range part of the strong force is mediated by the lightest meson:  the
pion. Figure (2a) illustrates the one-pion-exchange potential(OPEP), whose
iterates (in the Schr\"{o}dinger equation) dominate the potential energy, as we
will see.  This is also the origin of the bulk of the tensor force.  An
isospin-violating potential~\cite{pig} (of the same range) that incorporates
both a virtual $\gamma$ and $\pi$ is shown in Fig.\ (2c).  Part of the
two-pion-exchange potential (TPEP), due to the exchange of uncorrelated
(non-interacting) pions, is depicted in Fig.\ (2b), while the exchange of a
typical heavy meson (correlated or interacting pions) is portrayed in Fig.\ (2d).

\begin{figure}[htbp]
  \epsfig{file=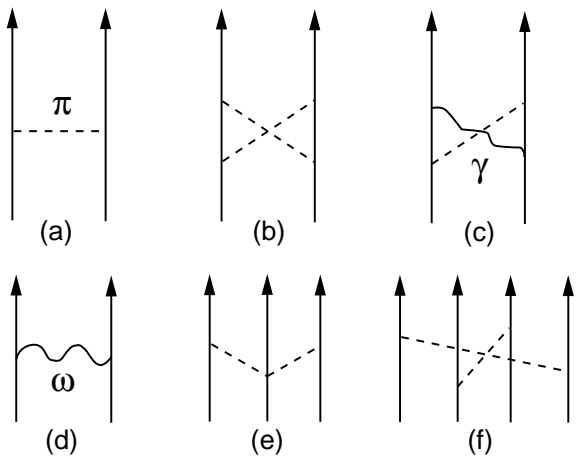,height=2.50in,bbllx=80pt,bblly=393pt,bburx=464pt,
  bbury=640pt}
  \caption{Nuclear-force components, some of which (a,b,d,e) are represented 
  (at least schematically) in nuclear potential models. Dashed lines depict
   pions.}
\end{figure}

Figure (3) schematically illustrates the result.  The tail of $V(r)$ is OPEP
(plus the much smaller $\pi$-$\gamma$-exchange potential).  As we move inward
TPEP plays a role, followed by heavy-meson exchange and the shortest-range
cutoff.  The short-range parts have always been problematic.  All potentials
contain OPEP, while most contain some form of two-pion exchange (although not
the asymptotically correct one).  Some potentials, such as the $AV_{18}$, have a
purely phenomenological interior, while others incorporate known heavy mesons
and other physics, yet both types are effective.  How can this be so?

\begin{figure}[htbp]
  \epsfig{file=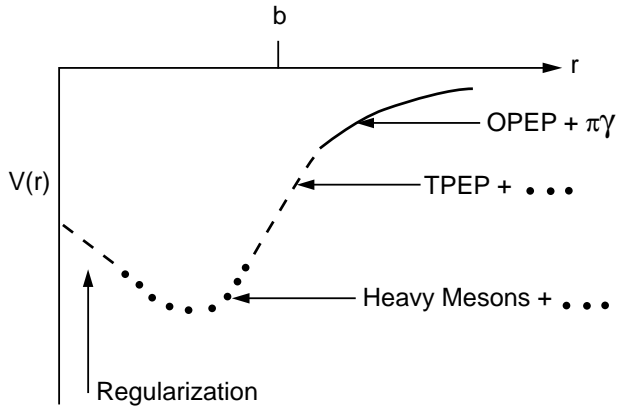,height=2.20in,bbllx=35pt,bblly=347pt,bburx=478pt,
   bbury=580pt}
  \caption{Cartoon of nuclear potential, showing the various regions of 
  importance.}
\end{figure}

Nuclei are basically low-momentum systems.  Consequently, short-range parts of
potentials are not probed in detail.  In essence the situation is like an
effective-range expansion:  only a few moments of the NN potential are actually
needed in almost all applications.  Thus any form of parameterization works,
whether it is pure phenomenology or a detailed treatment of the physics.  This
synopsis of the situation (illustrated later) also reflects the philosophy of
$\chi$PT:  the short-range interactions are counter terms fit to data.  If the
organization of the problem is effective, only a few such counter terms are
required for each spin-isospin component.

The innermost region does more than determine moments of V; it regulates the
divergences present in any potential.  Nuclear mechanisms such as OPE typically
fall off slowly in momentum space (or not at all), and this leads to divergent
loops when the Schr\"{o}dinger equation is solved.  Providing support in
momentum space (smoothing short-distance behavior in configuration space) is the
function of these regulators (usually called form factors).  Fitted form factors
have a typical range of $\Lambda \sim 1$ GeV, which is probably not an accident
given the range of convergence of the chiral expansion~\cite{jim}. Having
regulated the loops, the fitting procedure then provides a kind of
``renormalization'' as well, if we view renormalization as a consistent scheme
for relating different experimental observables.  These observables are
appropriate to the $\sim$ 1 GeV scale of the regulators.

However fundamental this description of nuclear-potential methodology is viewed,
the black box is very well tuned.  Most few-nucleon quantities that are
calculated (and there are many) agree well with experiment.  I will generally
ignore these successes (see, however, the impressive achievements reported in
Refs. ~\cite{joe,walter}). The few problems that exist are very interesting and
appear to be correlated.

Figure (3) also shows the characteristic radius, b (=1.4 fm),
used in the Nijmegen PSA~\cite{psa}. Outside that distance OPEP and long-range
electromagnetic interactions are incorporated.  Inside that distance the
treatment is purely phenomenological.  One can also relax this separation and
fit the entire potential, incorporating as much (or as little) physics in the
interior region as desired.  The best of the Nijmegen-fitted second-generation
potentials have the ($\chi^2$-determined) quality of their best PSA.  These 
ideas are
illustrated~\cite{psa} in Fig. (4), which depicts the $^3$P$_0$ NN phase shift
(in degrees) as a function of lab-frame NN kinetic energy.  The dashed
curve is the result of no interaction inside b = 1.4 fm and OPEP (plus some
additional two-pion-range interaction) outside.  This seemingly modest physics
input reproduces the shape of the phase shift but not the position of its zero. 
Incorporating an inner region specified by a single parameter leads to the
dotted curve, which is an excellent fit in all but the finest details. 
Two additional parameters produce the ultimate solid curve, which perfectly
matches the single-energy phase-shift points with error bars.  We note, however,
that the Nijmegen PSA is a multi-energy phase-shift analysis (data at all
energies are simultaneously fit), which is a superior procedure and the strength
of the method~\cite{vanc,psa}. For all but the most sophisticated treatments, a
single short-range parameter suffices in the $^3$P$_0$ partial wave.

\begin{figure}[htbp]
  \epsfig{file=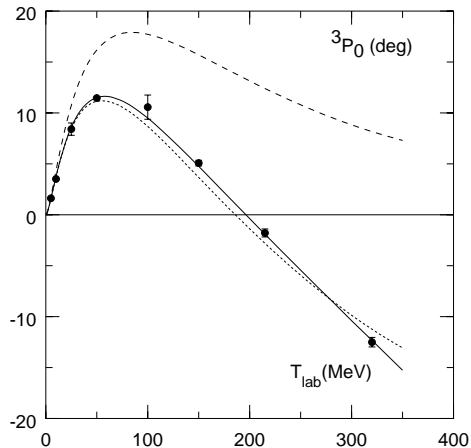,height=2.50in,bbllx=-70pt,bblly=250pt,bburx=650pt,
   bbury=650pt}
  \caption{$^3P_0$ phase shift calculated with OPEP tail for $r > b$ (dashed 
  line), and with one (dotted) and 3 (solid) short-range-interaction parameters 
  added.}
\end{figure}

The physics in the tail of the NN force can be quantitatively tested in a PSA.
The Nijmegen PSA~\cite{psa,vanc} determines the strength of the various 
$\pi$-N-N coupling constants ($\pi^0 pp$, $\pi^0 nn$, and $\pi^{\pm}np$) while
(as a check) determining the $\pi^0$ and $\pi^\pm$ masses that parameterize OPEP
(i.e., in the combination $V_{\pi}(m_{\pi}\, r)$).  They find that within
$\sim$1\% errors the 3 coupling-constant combinations are the same, and if they
are assumed to be identical their latest results~\cite{vanc} lead to the 
composite value
$$
f^2 = 0.0749(3)\, . \eqno (1)
$$
This is equivalent to $G = 13.05(2)$ and $G^2/4 \pi = 13.56(5)$, with the
Goldberger-Treiman relation (written as an identity using the discrepancy,
$\delta_{GT}$)
$$
\frac{G}{M} (1 - \delta_{GT}) = \frac{g_A}{f_{\pi}}\, , \eqno (2a)
$$
producing
$$
\delta_{GT} = 1.9(4)\%\, , \eqno (2b)
$$
which has a natural size ($\sim m^2_{\pi} / \Lambda^2$).  The Nijmegen analysis
also finds\cite{mass}
$$\eqalignno{
m_{\pi^\pm} &= 139.4(10)\, {\rm MeV}\, , &(3a)\cr
m_{\pi^{0}} &= 135.6(13)\, {\rm MeV}\, . &(3b)\cr}
$$
The very small errors on $m_{\pi}$ underscore the importance of  
OPEP in the nuclear force.  In the less-well-measured reaction $p  
\bar{p} \rightarrow \Lambda \bar{\Lambda}$, the same type of analysis 
led~\cite{leap} to $m_K = 475 (30)$ MeV, which agrees with measured masses.
Equation (1) is also consistent with recent~\cite{pin} $\pi N$ PSA's.  Given 
this remarkable sensitivity to OPEP, can other components of the nuclear  
force be determined, as well?

It might be viewed as unremarkable that OPEP can be shown to be an  
important part of the nuclear force; it has to be there and its size is known. 
The TPEP is another matter entirely.  This potential has had a checkered 
history~\cite{twopi},  
because it is the first part of the nuclear potential that is affected in
leading order by off-shell choices and the first part that is strongly model  
dependent.  The model dependence is obvious:  chiral symmetry plays  
an essential role in weakening this force~\cite{bira2pi}.

The off-shell dependence is a very old story, but one that continues to be 
retold (including at this workshop~\cite{evgeni}).  The physics is very 
simple: a potential is a subamplitude meant to be iterated in the 
Schr\"{o}dinger  
equation.  As such, it is not an observable (i.e., it is unphysical) and  
different definitions (viz., different {\bf off-shell}  
subamplitudes) are possible that lead to identical on-shell amplitudes (which  
are observables).  Thus, it makes little difference which definition is  
used, as long as the approach is consistent.

I am aware of three forms of off-shell choices for TPEP at  
the subleading-order level.  The most important distinction is the  
Brueckner-Watson~\cite{bw} (BW) vs.\ Taketani-Machida-Ohnuma~\cite{tmo} (TMO)  
approach.  Schr\"{o}dinger (old-fashioned) perturbation theory
leads naturally to an energy-dependent potential, because energy  
denominators (from loops, etc.) contain the overall system energy [i.e.,
($E - H)^{-1}$], while energy transfers $(q_0)$ are contained in pion  
propagators: $(\mbox{\boldmath $q$}^2 + m^2_{\pi} - q^2_0)^{-1}$.   
Retaining the energy dependence in OPEP (implicitly) was the BW approach, while
eliminating it was the TMO approach.  Long ago a variety of formal  
approaches were developed to deal with removal of energy dependence  
(reviewed for the deuteron problem in Ref. \cite{quasi}).  Each has its strong  
and weak points.  The choice of BW or TMO leads to a different TPEP  
at leading order (see Ref. \cite{twopi} for a detailed discussion), although a 
very simple transformation relates the two.  At subleading order, unitary  
equivalences and ``form'' equivalences~\cite{quasi} (viz., using $H^2 \Psi = 
E^2 \Psi$, rather than $H \Psi = E \Psi$) enter.  We emphasize that none of  
these approaches is inherently correct or incorrect.  Applications of a  
particular approach will be either correct or incorrect. I personally prefer
the TMO approach because it makes the nuclear many-body problem much more 
tractable~\cite{3nf}.

The first consistent calculation of TPEP using a chiral Lagrangian was  
performed by Ord\'o\~nez and van Kolck~\cite{bira2pi}, who developed the 
BW form of the potential (the TMO form was given later in Ref. \cite{twopi}).  
This result has been recently  
confirmed by the Nijmegen group~\cite{rob} (whose methods require the TMO
form), who added this potential to OPEP for r $>$ b.  As a check of their  
procedure, they verified that TPEP is a function of $(2  m_{\pi}\, r)$, and 
that the strength is given by $f^4$.  This is an important  
result that directly tests the chiral mechanism in nuclear forces. Not only 
does this force significantly improve the overall fits to the NN data (compared 
to using OPEP alone in the tail of the force), but it will facilitate 
the construction of ``third-generation'' potentials~\cite{johan} that contain 
improved physics (TPEP plus the $\pi$-$\gamma$ force).  Finally, although we 
know of no direct evidence for heavy-meson exchange, this mechanism is 
universally accepted.

\section{The Few-Nucleon Problem}

The full machinery of the GFMC method~\cite{a6-8} was unleashed on light nuclei 
by the Urbana-Los Alamos-Argonne collaboration.  Their Hamiltonian  
contained the AV$_{18}$ potential (incorporating OPEP, a schematic form of  
TPEP, and a purely phenomenological short-range part) as well as the Urbana  
IX three-nucleon force [3NF] (which contains a schematic form of  
two-pion-exchange 3NF [such as Fig.\ (2e)], but no 4NF, such as Fig.\ (2f)).   
Their results can be summarized as follows.  The deuteron ($^2$H)  
and triton ($^3$H) bound states are fit to the NN and 3NF models.   
The alpha particle ($^4$He) is a prediction and agrees with  
experiment.  The required 3NF is weak.  The nucleus $^5$He is actually  
two $n - \alpha$ resonances ($p_{1/2}$ and $p_{3/2}$) and their  
calculated splitting is 20-30\% too small, but the agreement with experiment is 
otherwise good.  There is underbinding in A = 6, 7, and 8 nuclei that is 
isospin dependent (higher-isospin states are worse).  The experimental  
binding-energy difference of the isodoublet $^3$He and $^3$H (764 keV) is in 
good agreement~\cite{AV18} with calculations (750(25) keV).  The too-small 
spin-orbit  
splitting is identical to the so-called ``$A_y$ puzzle'', where the nucleon  
asymmetry in N-d scattering at low energy is too small by comparable amounts.  
A recent study~\cite{dirk} of the latter suggests that no modification of  
the NN force allowed by NN data can account for the puzzle.

It therefore appears plausible that a better 3NF containing more physics should
be employed.  Such forces will contain operators that have a spin-orbit
character (one such term~\cite{3nf} is required by Lorentz invariance), although
only detailed calculations will determine whether magnitudes (or even signs) of
such terms are appropriate to resolve the problems.

Finally, one can use these calculations to estimate the size of various  
contributions to the energy.  Using results for A = 2, 3, and 4, one finds
$$\eqalignno{
\langle V_{\pi}    \rangle / \langle V \rangle &\sim \quad  70-80\%\, , &(4a)\cr
\langle V_{\pi}          \rangle &\sim -15 \,{\rm MeV/ pair}\, , &(4b)\cr
\langle V_{\rm sr} \rangle       &\sim -5  \,{\rm MeV/ pair}\, , &(4c)\cr
\langle V_{3NF} \rangle        &\sim -1 \,{\rm MeV/ triplet}\, , &(4d)\cr
\langle T  \rangle             &\sim \ 15 \,{\rm MeV/ nucleon}\, , &(4e)\cr
\langle V_{\rm C} \rangle &\sim \ \; \frac{2}{3} \,{\rm MeV/ pp \; pair}\, . 
&(4f)\cr}
$$
The quantities $V_\pi, V_{\rm sr}, V_{3NF}, V_{\rm C}$, and $T$ are the OPEP, 
short-range, 3NF, Coulomb, and kinetic-energy parts of the Hamiltonian.
Our final task is to see whether these sizes can be understood in terms  
of nuclear scales.

\section{Chiral Scales}

The building blocks of $\chi P T$ are pion fields, derivatives, and  
nucleon fields coupled to form zero-range interactions.  According to  
Manohar and Georgi~\cite{ndpc}, the generic Lagrangian can be written  
schematically as
$$
\Delta {\cal{L}} = c_{\ell m n} \left(\frac{\bar{N} (\cdots) N}{f^2_{\pi}  
\Lambda} \right)^{\ell} \left(\frac{\mbox{\boldmath  
$\pi$}}{f_{\pi}}\right)^m \left( \frac{\partial^{\mu}, m_{\pi}}{\Lambda}  
\right)^n f^2_{\pi} \Lambda^2\, , \eqno (5)
$$
where $\Delta = n + \ell - 2 \geq$ 0 is the chiral constraint that makes  
$\chi P T$ work, $f_{\pi} = 92.4$ MeV, $\Lambda \sim 1$ GeV (as  
before), and $c_{\ell m n} \sim 1$.  The latter condition,  
``naturalness'', makes it possible to check scales in a convincing  
manner.  Various matrices can be inserted between the nucleon fields.   
Note that the dependence on $\Lambda$ is given by $\Lambda ^{-  
\Delta}$.  Moreover, if we set $m = n = 0$ and look at the resulting  
class of zero-range interactions $[(\bar{N} N)^{\ell}]$, we see that  
$\ell = 2$ corresponds to a two-body interaction, $\ell = 3$ to a  
three-body interaction, $\ell = 4$ to a four-body interaction, etc., and they 
behave as $\Lambda^0, \Lambda^{-1}, \Lambda^{-2}, \ldots$, respectively.  
Since $\bar{N} N$ is like a nuclear density, and the density of nuclear matter 
is $\sim 1.5 f_{\pi}^3$, the relative size~\cite{scales} of these many-body 
operators is $\sim \left( \frac{1.5  
f_{\pi}}{\Lambda} \right)^{\ell} \sim \left( \frac{1}{7} \right)^{\ell}$  
in normal nuclei (or smaller), and this converges fairly rapidly.

Is there a way to be more quantitative?  Nuclear physics is not usually
formulated in terms of zero-range forces.  However, in 1992 a
group at Los Alamos~\cite{dave} performed relativistic-Hartree-approximation  
calculations using zero-range forces of the same generic form as  
Eq.(5), but with no pions.  Their motivation was to find an easy way to  
do Hartree-Fock calculations, where two nucleons are exchanged in an  
interaction.  For zero-range interactions this can be rewritten using  
Fierz identities as a linear combination of the original Hartree terms.   
Thus, a single calculation is at the same time both Hartree and Hartree-Fock.
Since the Fierz coefficients are numbers on the order of 1, if the Hartree  
coefficients $c_{\ell m n}$ are natural, the Hartree-Fock results  
(simply a rearrangement for each class of Hartree term) will also be natural.

\begin{figure}[htbp]
  \epsfig{file=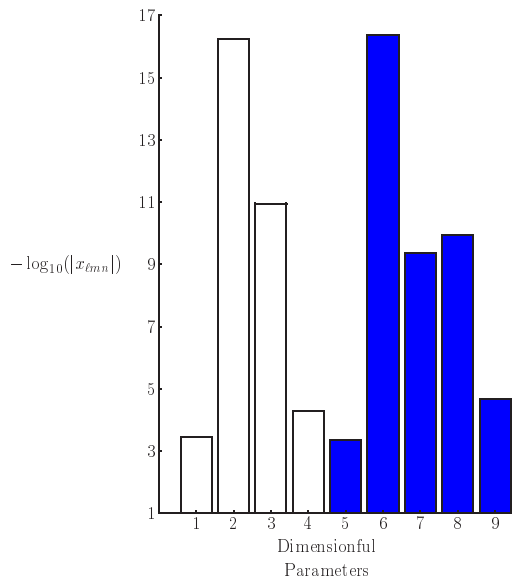,height=2.750in,bbllx=-150pt,bblly=170pt,bburx=600pt,
   bbury=650pt}
  \caption{Negative logarithm of dimensionful parameters, $|x_{\ell m n}|$.
  Negative values of $x_{\ell m n}$ are shaded.}
\end{figure}

The calculation of Ref. \cite{dave} fit the properties of 3 nuclei with 9  
parameters [4 types of $c_{200}$'s, $c_{300}$, two $c_{400}$'s,  
and two $c_{202}$'s] and predicted (quite accurately) the properties  
of 57 other nuclei.  A large number of possible operators of orders  
$\Lambda^0, \Lambda^{-1}$, and $\Lambda^{-2}$ were not included (the  
authors were not trying to perform a chiral expansion), but the exercise  
was very successful.  During Bryan Lynn's visit to Los Alamos in 1994, 
we~\cite{bryan} realized that the coupling constants of that calculation could 
be checked for naturalness.  The results delighted us.
Although their coupling constants had been fitted using a $\chi^2$  
procedure, their published values (used by Bryan and me) were, however, not  
the minimum-$\chi^2$ solution, which seemed to them not to be as  
predictive as the one chosen.  The best-$\chi^2$ solution is shown  
graphically in Figs.\ (5) and (6).  Figure (5) shows the coefficients  
(called $|x_{\ell m n}|$) with the dimensional factors of $f_{\pi}$ and  
$\Lambda$ (in units of MeV$^{-n}$) included.  The 5 negative  
\begin{figure}[th]
  \epsfig{file=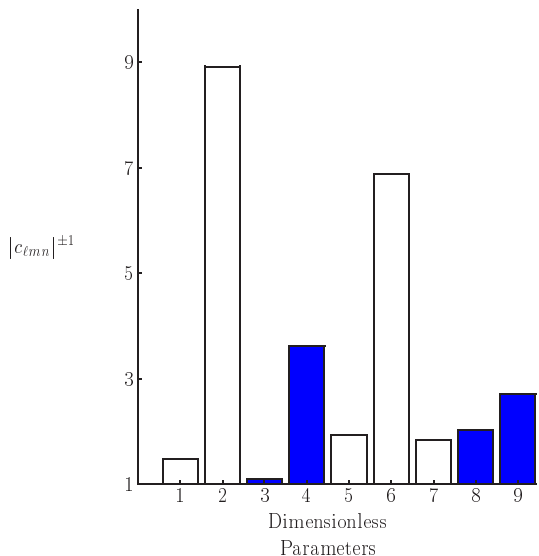,height=2.750in,bbllx=-100pt,bblly=250pt,bburx=500pt,
   bbury=675pt}
  \caption{Magnitude of dimensionless parameters $|c_{\ell m n}|$ (if greater 
  than 1), and 1/$|c_{\ell m n}|$ (shaded, if less than 1).}
\end{figure}
coefficients are ordered to the right.  The coefficients group according  
to the dimensionful factors (e.g., the lowest values are the 4  
$c_{200}$'s), which vary over 14 orders of magnitude.  Figure (6)  
shows the dimensionless parameters grouped in the same way as in  
Fig.\ (5). If a $|c_{\ell m n}|$ is greater than one (in magnitude) it is 
plotted with no shading, while if it is less than one, $1/|c_{\ell m n}|$ is 
plotted instead and the result is shaded. Most 
of the values are small and natural by anyone's criterion.  The two  
large values (one positive and one negative) multiply operators with  
the same nonrelativistic limit, and in that limit the effective coefficient 
(the sum of the two) is natural (2.0), while the difference (corresponding in 
a nonrelativistic expansion to higher-order (in $1/ \Lambda$) operators) is  
very large.  I suspect that this latter result reflects missing operators in  
the chiral expansion, but we may never know.  Better calculations by Furnstahl 
and Rusnak~\cite{pcnm} have reached the same conclusion:  nuclei reflect
fundamental scales and display naturalness.  As an aside, we note that  
the coupling constants of heavy bosons (to nucleons) exchanged between two
nucleons can be shown~\cite{scales} to be $\sim \Lambda / f_{\pi} \sim 10$ if 
they are  
natural.  Indeed, this is the generic size of all such coupling constants.

\section{Nuclear Scales}

Having remarked earlier that pions play an exceptional role in nuclei, it  
is necessary to reintroduce them into the power-counting scheme.   
First, however, we need to set the momentum scale inside nuclei.  This  
scale can be set~\cite{scales} by an inverse correlation length (R), by an  
uncertainty-principle argument, or by examining Fig.\ (7), which
illustrates energy accrual in the triton. If one fixes the distance between 
any two nucleons to be $x$, one can easily calculate the potential or kinetic
energy accruing inside that distance (which requires integrating over the 
coordinates of the third nucleon).
As $x \rightarrow \infty$, clearly all of the energy is calculated, so that  
after dividing the accrual inside $x$ by the total, the percentage accrual can 
be plotted, as in Fig.\ (7). As $x$ increases from the origin, the short-range  
repulsion is evident (the net potential energy is attractive, so negative  
values on this plot are repulsive).  Most of the action takes place  
between 1 and 2 fm, corresponding to 100-200 MeV/c, and we choose to  
take Q (the characteristic momentum) on the order of a pion mass (a  
convenient mnemonic):
$$
Q \sim 1/R \sim m_{\pi}\, . \eqno (6)
$$
We note that $Q$ is about half the Fermi momentum, $k_F$, which  
sets a scale in nuclear matter. We also note that the kinetic and potential 
energies closely track each other.

\begin{figure}[htb]
\epsfig{file=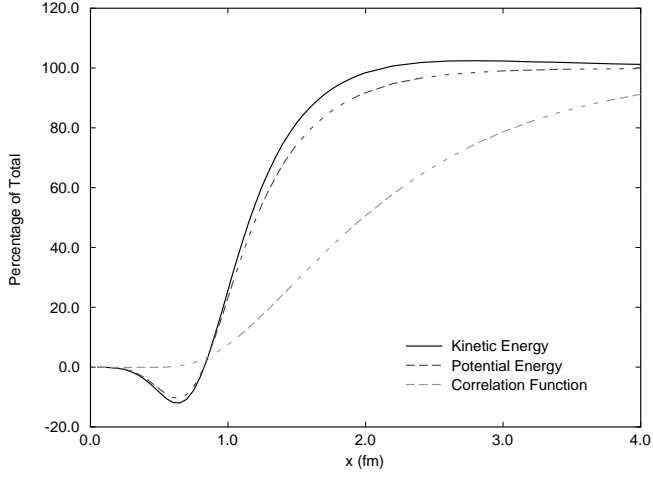,height=3.0in,bbllx=-20pt,bblly=400pt,bburx=593pt,
 bbury=773pt}
\caption{Percentages of accrual of kinetic energy (solid line),
potential energy (short dashed line), and probability (long dashed line)
within an interparticle separation, $x$, for any pair of nucleons in the
triton.}
\end{figure}

In terms of this correlation length or typical momentum, we can  
estimate the expectation values in a nucleus for various energy operators 
by rewriting their well-known expressions~\cite{scales} in terms of $Q,  
f_{\pi}$, and $\Lambda$.  Doing this we find
$$\eqalignno{
\langle V_{\pi} \rangle &\sim \frac{Q^3}{f_\pi \Lambda} \sim 25  
  \, {\rm MeV/ pair}    \, , &(7a)\cr
\langle T \rangle       &\sim \frac{Q^2}{\Lambda}  \    \sim 20
  \, {\rm MeV/ nucleon} \, , &(7b)\cr
\langle V_{\rm sr} \rangle &\sim \frac{Q^{3}}{f_{\pi} \Lambda}, &(7c)\cr
\langle TPEP \rangle &\sim \frac{Q^5}{f_\pi \Lambda^3} \: \sim  
  \frac{3}{4} \, {\rm MeV/ pair}\, , &(7d)\cr
\langle V_{3NF} \rangle &\sim \frac{Q^6}{f^2_{\pi} \Lambda^3} \:
  \sim \frac{1}{2} \, {\rm MeV/ triplet}\, , &(7e)\cr
\langle V_{\rm C} \rangle &\sim \ \; \alpha Q \ \; \sim \, 1\, {\rm MeV/ 
pp\; pair}\, . &(7f)\cr}
$$
This exercise has been conducted in configuration space, and we have  
used $\Lambda \sim 4 \pi f_{\pi}$.  Some creativity was needed for the  
short-range potential, $V_{\rm sr}$, and (7c) is more like an upper limit.   
The factor of $Q^3$ in (7a) and (7c) is easy to understand, since it just
reflects the phase-space factor needed to convert a momentum-space  
expression ($\sim Q^0$) to configuration space, while a residual $1/4  
\pi$ from that exercise produces the $1/ \Lambda$ (from $4 \pi  
f_{\pi}$).  We note that these results are consistent with what we  
inferred earlier from the GFMC calculations of A = 2-4 nuclei.

A more systematic treatment was developed in Ref. \cite{scales}, which leads to
$$\eqalignno{
\langle \hat{E}_{\rm irr} \rangle &\sim \frac{Q^{\nu}}{f^{n_{\rm c}}_{\pi}
\Lambda^{2L + n_{\rm c} + \Delta}}\, , &(8a)\cr
\nu &= 1 + 2 (L + n_{\rm c}) + \Delta\, , &(8b)\cr
\Delta &= \sum_{i} (d_i + f_i/2 -2)\; \geq 0\, , &(8c)\cr}
$$
where $L$ is the number of loops, and $n_{\rm c}$ is a topological parameter 
that equals the number of interacting nucleons minus the number of clusters  
(of interacting nucleons).  The quantity $\Delta$ is basically ($n + \ell -  
2$) from Eq. (5) summed over all vertices used to construct the irreducible 
energy operator $\hat{E}_{\rm irr}$.  This neat formula summarizes the previous 
ones, and is equivalent to Weinberg's result~\cite{cpt} ($Q^\nu$) with factors 
of $f_{\pi}$ and $\Lambda$ put in.

We note that leading-order calculations ($\Delta = 0, L = 0$) depend  
only on $n_{\rm c}$, whose smallest value is $N - 1$ for $N$ interacting  
nucleons.  $N$-body forces in a nucleus therefore scale as ($Q/  
\Lambda)^{N-1}$ and decrease as $N$ increases.  This is also a  
minimal condition for nuclear tractability, since very large $N$-body  
forces would nullify our calculational ability in nuclei. Each new nucleus
(as we increase $N$) would require large, new, and very complicated forces for
its description.

Naive applications of Weinberg power counting were shown to be ineffective in
several talks~\cite{bira,david,martin} at this workshop, and yet we have seen
that nuclei exhibit power counting.  How can both be true?  It is in
treating the reducible graphs that the original power counting has difficulty.
The reducible graphs are necessary in order to build nuclear wave functions (in
the language of traditional nuclear physics) obtained from iterating the
Schr\"{o}dinger equation.  Equation (8a) summarizes and uses that wave-function
information (however obtained) in calculating the expectation value of an
irreducible operator.  Hence the two seemingly contradictory statements are not
in disagreement.  The power counting we have seen here refers to how various
$N$-body building blocks work inside a nucleus. This was called a ``hybrid''
approach earlier. Summing reducible graphs with a credible power-counting scheme
is much more difficult, as we have seen at this workshop.

\section{Summary}

In summary, we have shown that OPEP dominates in few-nucleon  
systems.  It has been ``measured'' in the Nijmegen PSA program.   
Very recently TPEP has been similarly ``measured'', and the chiral form of 
this force has been verified.  We have argued  
that the short-range parts of the nuclear potential enter the physics at  
low energy in the form of moments, only a few of which optimally  
incorporate the constraints of the $NN$ data for any spin-isospin  
combination in the nuclear force.  Most calculated few-nucleon  
observables agree very well with the experimental data, with a notable  
exception being spin-orbit-type observables, such as $A_y$ in $N-d$  
scattering and the $p_{1/2} - p_{3/2}$ splitting in $^5$He.  There is  
also a problem with higher-isospin states for $A \geq 6$.  Isospin  
violation in $^3$He - $^3$H is well understood and is consistent with  
natural mechanisms.  Finally, three-nucleon forces are weak  
($\sim$1\%) but necessary when used with conventional $NN$ interactions.  The  
weakening of $N$-body forces is a necessary condition for nuclear  
(calculational) tractability. 

\section*{Acknowledgements}

The work of J.L.F. was performed under the auspices of the U.S. Department 
of Energy. The help of J.J. de Swart and R.G.E. Timmermans is gratefully 
acknowledged.

\end{document}